\documentclass{aa}            
\usepackage{times,mathptm}
\usepackage{epsfig}

\newcommand{\kms}{~km~s$^{-1}$~}
\newcommand{\plm}{\mbox{$\pm$}}
\newcommand{\mua}{\mbox{$\mu_{\alpha}\cos\delta$}}

\newcommand{\mud}{\mbox{$\mu_{\delta}$}}

\def\tabvd{\mbox{$\stackrel{\displaystyle .}{\displaystyle .}$}}

\begin{document}
   \thesaurus{4         
              (10.07.2
               10.11.1
               10.08.1
               10.05.1)
                       }

   \title  {A proper motion study of the globular cluster M10}
   \author {L.~Chen
           \inst{1,2}
      \and  M.~Geffert
           \inst{2}
      \and  J.J.~Wang
	   \inst{1}
      \and  K.~Reif$\,$
           \inst{2}
      \and  J.M.~Braun
           \inst{2}
             }

   \offprints{M.~Geffert, Bonn}

   \institute  {          
              Shanghai Observatory, Chinese Academy of Sciences, 
              Shanghai 200030, China
      \and
              Sternwarte der Universit\"at Bonn, Auf dem H\"ugel 71,
              D-53121 Bonn, FRG
    }

   \date{Received date; accepted date}

   \maketitle

   \begin{abstract}
We present the first proper motion study of M10 (NGC~6254).
Absolute proper motions of about 532 stars in the field of the globular cluster
M10 were determined with respect to Hipparcos and ACT reference stars.
In addition to photographic plates of Bonn and Shanghai also wide field
CCD observations as second epoch plates were used.
The wide field CCD observations show an accuracy comparable to
that of the photographic plates.
A good coincidence of the solutions based on reference stars
from Hipparcos and from ACT was found.
Our final proper motions allow a sufficient separation
of cluster and field stars.
Two population~II Cepheids were confirmed to be members of M10.
The absolute proper motion of M10 was determined and combined with
its distance from the Sun and its radial velocity.
The space motion and metallicity of M10 indicates the characteristics of 
a halo  object with an orbit reaching to a maximal $z$-distance of less
than 3~kpc. 

   \end{abstract}

\section{Introduction}

Globular clusters are important tracers of the chemical and dynamical
evolution of the Milky Way. Since the orbits of globular
clusters may have kept their characteristics from the
early times of the formation of the Milky Way, 
we may extract from their kinematics 
some clues for the understanding of the origin of the Milky Way.
However so far for only some 40 globular clusters 
absolute proper motions exist (Dinescu et al.\, 1999), 
which partly are of low quality
(e.g. Dauphole et al.\, 1996). Nevertheless on the basis of
a  sample of 26 globular clusters
Dauphole et al.\, (1996) found an indication of a 
metallicity gradient among the halo clusters and a mean rotation 
of about +40 \kms ~for the complete globular cluster sample.
Additional data especial for southern clusters were determined
by Dinescu et al.\, (1999 and references herein).

The results of the Hipparcos mission offer new possibilities
for the determination of absolute proper motions of globular
clusters and other objects of interest for galactic kinematics.
Since previous studies have used mainly extragalactic objects 
as proper motion zero points, the use of Hipparcos reference
stars offers a unique possibility for the determination 
of absolute proper motions at low galactic latitutes without
suitable extragalactic background objects.
Geffert et al.\, (1997) have determined absolute proper motions
of 10 globular clusters with respect to Hipparcos. These were 
combined with data of 5 other clusters and the kinematics of 
this sample was studied by Brosche et al.\, (1997) and 
Odenkirchen et al.\, (1997). A complete discussion of these
data together with additional proper motions was
given recently in Dinescu et al.\, (1999). 
 
M10 is a globular cluster located very near to the direction to the
galactic center ($l=15^\circ$, $b=23^\circ$), but at a distance of only
4.3~kpc from the Sun. Due to its
metallicity of $[\mbox{Fe/H}] = -1.52\;$dex (Harris 1996) it belongs to
the halo group of clusters according to Zinn's (1985) classification. 
The preliminary kinematical data of M10 (Geffert et al.\, 1997) indicate
a disk like motion, the most disk like motion found in the group of
globular clusters, whose proper motions were determined using Hipparcos
(Odenkirchen et al.\, 1997).
However, this preliminary proper motion has the problem that only three to
four Hipparcos stars could be used for the reduction of the plates.
This made the previous solution very uncertain.
Here we have included new photographic material (from Shanghai) and recent
CCD observations to get a new determination of the absolute proper motion
of M10. 
Moreover, we use in addition to the Hipparcos catalogue the ACT catalogue
(Urban et al.\, 1998).
The ACT catalogue, due to its denser coverage of the sky, allows the use
of more reference stars for the determination of the absolute proper motions.

\begin{table}
\caption[]{Photographic material used in this work}
\begin{flushleft}
\begin{tabular}{lrrr}
\hline\noalign{\smallskip}
  Plate No.  & Epoch  & Emulsion
  & Telescope  \\
\noalign{\smallskip}
\hline\noalign{\smallskip}
  149     & 1902.6 &             & DR HL \\
  406     & 1905.6 &             & R SH \\
  262     & 1916.5 &             & DR HL \\
  309     & 1917.5 &             & DR HL \\
  312     & 1917.5 &             & DR HL \\
  314     & 1917.5 &             & DR HL \\
 CL57016  & 1957.5 &  103a-O     & R SH \\
 CL57031  & 1957.6 & 103a-O      & R SH \\
 CL57033  & 1957.6 & 103a-O      & R SH \\
 1232     & 1976.6 & IIa-O/BG 25 & DR HL \\
 1874     & 1994.5 & IIa-O/BG 25 & DR HL \\
 1875     & 1994.5 & IIa-O/BG 25 & DR HL \\
\noalign{\smallskip}
\hline
\multicolumn{4}{l}{\scriptsize DR HL = ($D$=0.3~m, $f$=5.1~m) double refractor
of Bonn} \\[-0.05cm]
\multicolumn{4}{l}{\scriptsize R SH = ($D$=0.4~m, $f$=7.0~m) refractor of
Shanghai observatory}
\end{tabular}
\end{flushleft}
\label{t_photmat}
\end{table}

\section {Observations and the reduction}

Table~\ref{t_photmat}  lists the photographic plates used in this work,
while Table~\ref{t_ccdobs}  indicates the observational data of the CCD frames.
The plates were taken either with the ($D=0.3\;$m, $f=5.1\;$m)
double refractor of Bonn, now located at Hoher List observatory,
or with the ($D=0.4\;$m, $f=7\;$m) refractor of Shanghai. 
The CCD frames were taken with the WWFPP camera 
(Reif et al.\, 1994) at the 1.23~m telescope at Calar Alto 
and with a similar camera (HoLiCam) at Hoher List 
observatory.
The use of the complete material allows a nearly uniform coverage of 
the epoch difference of 92~years.
The limiting magnitude of the plates is of the order
of $V = 15.5\;$mag, while the corresponding one of the CCD frames
ranges from $V = 16$ to 19~mag.
The refractor plates of Bonn were mainly scanned at the
PDS~2020GM of University of M\"unster.
The plates R1874 and R1875 with lower limiting magnitude have been 
measured at the ASCORECORD of Hoher List observatory.
On these plates only 80 stars for each plate were measured.
Stars were extracted and rectangular coordinates $x$ and $y$ were
determined from the PDS measurements using standard 
procedures (e.g. Tucholke 1994).
The first epoch plates of the Bonn refractor contained scratches
and reseau lines, which led to problems for a significant
number of the stars.
Therefore for some of these stars no rectangular coordinates could be obtained.
The plates from Shanghai were scanned at the PDS 1010MS
of the Purple Mountain Observatory, Chinese Academy of sciences
(see also Wang et al.\, 1999).

The determination of the rectangular coordinates $x$, $y$ of the stars
on the CCD frames was performed for the observations from Hoher List by
standard CCD reduction techniques (DAOPHOT, IRAF) routines.
Magnitudes, $x$ and $y$ positions were determined via PSF fit.
The observations from Calar Alto were reduced earlier (Geffert et al.\, 1994)
by the IMEX routine of the IRAF program package.

The astrometric reduction was performed by a central overlap algorithm.
Due to the small field of the CCD frames we had to use only the plates
in the first step of the reduction.
A catalogue of positions and proper motions of 450 stars was established
in this first step.
In the following steps of the reduction the CCD frames were included.
While for the plates only quadratic polynomials of the rectangular coordinates
had to be taken into account, third order polynomials were necessary for the
reduction of the CCD frames.
The third order polynomials for the reduction of the CCD frames
had to be used due to the distortion of 
the optics of the focal reducer of the WWFPP
camera (Geffert et al.\, 1994).
From the different position and time pairs we determined for each star
for a certain epoch the mean position and the proper motion using
least squares technique.
All stars with proper motion errors larger than 4~mas/yr were omitted.
The final catalogue contains 532 positions and proper motions of stars
in the region of M10.
The median of the internal errors was about $\plm 1\;$mas/yr. 
From a plot of the proper motions versus magnitude no 
magnitude equation was found in our data.
\begin{table}
\caption[]{CCD observations used in this work.
           All CCD observations were made with the WWFPP camera
           (Reif et al. 1994) }
\begin{flushleft}     
\begin{tabular}{lclr}
\hline\noalign{\smallskip}
Telescope & Epoch  & Filter & No. of frames \\
\noalign{\smallskip}
\hline\noalign{\smallskip}
 1~m Hoher List    & 1996 & $V$, $B$ & 10 \\
 1.23~m Calar Alto & 1994 & $R$      &  5 \\
\noalign{\smallskip}
\hline
\end{tabular}
\end{flushleft}
\label{t_ccdobs}
\end{table}

\begin{table*}
\parbox[c]{4.5cm}{
\caption[]{The catalogue of positions and proper motions of 532~stars
in the field of M10.
Four lines are given as an example.
The complete catalogue is available in electronic form at VizieR (CDS;
see Ochsenbein et al. 2000).
The epoch of the positions is 1950}
\label{t_cdsdat}
} \hfill
\begin{tabular}{lr@{ }r@{ }r@{}lr@{ }r@{ }r@{}lrrrrrrr}
\hline\noalign{\smallskip}
No. & \multicolumn{4}{c}{$\alpha_{2000}$} & \multicolumn{4}{c}{$\delta_{2000}$}
 & \multicolumn{1}{c}{$\sigma_{\alpha}$} & \multicolumn{1}{c}{$\sigma_{\delta}$}
 & \multicolumn{1}{c}{$\mua$} & \multicolumn{1}{c}{$\mud$} &
 \multicolumn{1}{c}{$\sigma_{\mua}$} & \multicolumn{1}{c}{$\sigma_{\mud}$} \\
 & [$^{\rm h}$ & $^{\rm m}$ & $^{\rm s}$ &] & [$^\circ$ & $'$ & $''$ &] &
\multicolumn{1}{c}{[$^{\rm s}$]} &  \multicolumn{1}{c}{[$''$]} &
\multicolumn{1}{c}{[mas/yr]} &  \multicolumn{1}{c}{[mas/yr]} &
\multicolumn{1}{c}{[mas/yr]} &  \multicolumn{1}{c}{[mas/yr]} \\
\noalign{\smallskip}
\hline\noalign{\smallskip}
1 & 16 & 54 & 34&.488 & $-4$ & 45 & 20&.57 &
  0.004 &  0.04 & $-33.2$ & $-27.5$ & 1.7 & 1.2 \\
2 & 16 & 54 & 46&.910 & $-3$ & 52 & 53&.73 &
  0.006 & 0.02 & 3.5 & 5.8 & 3.3 & 0.6 \\
3 & 16 & 54 & 47&.085 & $-4$ & 23 & 06&.77 &
  0.002 & 0.01 & $-5.5$ & 4.4 & 1.1 & 0.3 \\
4 & 16 & 54 & 48&.326 & $-4$ & 13 & 34&.70 &
  0.002 & 0.03 & $-4.3$ & $-2.2$ & 0.7 & 0.7 \\
$\;$\tabvd & \multicolumn{4}{c}{\tabvd} & \multicolumn{4}{c}{\tabvd} &
 \multicolumn{1}{c}{\tabvd} & \multicolumn{1}{c}{\tabvd} &
 \tabvd$\;\:\:$ & \tabvd$\;\:\:$ & \tabvd$\;\:\:$ & \tabvd$\;\:\:$\\
\noalign{\smallskip}
\hline
\end{tabular}
\end{table*}

We have performed two independent reductions of the M10 data with
reference stars from Hipparcos (ESA 1997) and the ACT catalogue
(Urban et al.\, 1998).
Although the Hipparcos stars proper motions are more accurate,
they seemed to be of only limited usefulness for our work,
since the majority of the plates contained only four Hipparcos stars,
while about 17 stars could be used from the ACT catalogue.
Therefore we consider both solutions as equivalent.
Table~\ref{t_cdsdat}  (the complete table is available in electronic form)
gives the catalogue of our positions and proper motions for the complete field
with respect to the ACT catalogue.
The limiting magnitude of this catalogue is about $V = 15.5\;$mag,
which corresponds to the limiting magnitude of the first epoch plates.
The size of the field is approximately $75\;\times\;75\;{\rm arcmin}^2$
centered on M10.
We chose the ACT solution for the catalogue in Table~\ref{t_cdsdat}.
The proper motions may be transfered to the Hipparcos system by adding
$\Delta\mua = -1.5\;$mas/yr and $\Delta\mud = +0.1\;$mas/yr
to the proper motions from Table~\ref{t_cdsdat}.
For the determination of the membership we will also use the solution
based on the ACT catalogue,
while for the determination of the absolute proper 
motion of M10 we will take the mean of both solutions.

\section {On the astrometric accuracy of the Hoher List CCD frames}

CCDs have been used in astrometry since several years,
e.g. for the determination of parallaxes, double stars and
for meridian circle observations (see references in Geffert 1998).
However most of these observations are based on CCD observations
with fields of $(10\arcmin)^2$ and smaller.
Since our study uses CCD observations of fields with a size at least 
20\arcmin$\;\times\;$20\arcmin\ it seems necessary to evaluate the accuracy,
which may be obtained with such telescope detector combinations.
While the CCD frames of Calar Alto were already tested in an earlier study
(Geffert et al.\, 1994),
we will concentrate here on the Hoher List observations.
 
In a first step we have compared positions of stars from pairs of CCD frames.
The positions of one frame were transformed by an affine transformation
to a second frame and the rms of the differences were calculated 
for each coordinate.
Under the assumption that both frames contribute with equal weight to the
differences, we have calculated from the rms the mean uncertainty of one
position on one frame.
These are given in Table~\ref{t_ccdacc} for several CCD pairs 
with nearly identical limiting magnitudes. 
In this comparison we included all stars which were detected
on the CCD frames.

\begin{table}
\caption[]{Mean accuracy of the position of one star on a CCD frame
           determined from intercomparison of two CCD frames}
\begin{flushleft}
\begin{tabular}{lrrrr}
\hline\noalign{\smallskip}
  Frames  & \multicolumn{1}{c}{$s_{\alpha}$} & \multicolumn{1}{c}{$s_{\delta}$}
  & Lim. mag. & No. of stars \\
  & [mas]  & [mas]  & \multicolumn{1}{c}{($V$)} &  \\
\noalign{\smallskip}
\hline\noalign{\smallskip}
4236/4237 &  62 &  64 & 16.5   & 1373 \\
4238/4239 & 120 & 120 & 18.5  & 3378 \\
4240/4241 &  63 &  67 & 16.2 & 1081 \\
4242/4243 & 100 &  80 & 17.0  & 1980 \\
\noalign{\smallskip}
\hline
\end{tabular}
\end{flushleft}
\label{t_ccdacc}
\end{table}

In a second test we consider only the stars which contribute
to the final catalogue of our investigation.
These are stars at the brighter end of the magnitude distribution.
Reducing the plates with one catalogue (ACT or Hipparcos)
leads to positions of the stars of each plate/CCD frame in a
common system.
The position and proper motion of each star in our final solution
described in Sect.~2 are determined by a fit to the positions and
epochs of the different plates/CCD frames for each star.
The mean position and proper motion of each star
is then used to update the positions of each star for the
epoch of the individual plates/CCD frames.
For each plate the mean and rms of the positional differences
from the initial positions are determined.
The rms will give an indication of the accuracy
of each  individual plate/CCD frame.
Table~\ref{t_posdiff}  summarizes for the Hoher List frames the mean and
standard deviations for each frame. 

\begin{table}
\caption[]{Mean deviation ($\Delta_{\alpha}$, $\Delta_{\delta}$) and
rms ($\sigma_{\alpha}$, $\sigma_{\delta}$) of the positions of
different CCD frames in the last step of the iteration.
The data are based on the mean of about~270 stars} 
\begin{flushleft}
\begin{tabular}{lrrrrr}
\hline\noalign{\smallskip} 
  CCD frame  & \multicolumn{1}{c}{$\Delta_\alpha$} &
  \multicolumn{1}{c}{$\sigma_\alpha$} & \multicolumn{1}{c}{$\Delta_\delta$} &
  \multicolumn{1}{c}{$\sigma_\delta$} & Colour \\
        & [mas]  &  [mas]  & [mas]  & [mas]  &     \\
\noalign{\smallskip}
\hline\noalign{\smallskip} 
 4236 & $+10$ &  70 & $-7$ & 90 & $B$ \\
 4237 &  $+6$ &  70 & $-5$ & 70 & $B$ \\
 4238 &  $+7$ &  80 &  $0$ & 90 & $B$ \\
 4239 &  $+6$ & 100 & $+2$ & 50 & $B$ \\
 4240 &  $+7$ &  80 &  $0$ & 50 & $V$ \\
 4241 &  $-2$ &  40 & $-1$ & 50 & $V$ \\
 4242 &  $-0$ &  90 & $+5$ & 50 & $V$ \\
 4243 &  $+3$ &  40 & $+6$ & 90 & $V$ \\
\noalign{\smallskip}
\hline
\end{tabular}
\end{flushleft}
\label{t_posdiff}
\end{table}

Table~\ref{t_posdiff}  shows a slight difference in the $B$ and $V$ frames
of the order of a few mas.
Nevertheless the systematic deviations between the $B$ and $V$ frames
are small. 
The positional accuracy of each star is of the order of one tenth of a pixel.
This value is a little bit larger with respect to other studies
(e.g. Geffert 1998). 
The reason may be the  crowding in the region of M10. 
In general,
the accuracy of one single frame is of the order of the accuracy of
one refractor plate, which justfies the use of CCD frames for the second
epoch observation.

\begin{figure}
\epsfysize=6.0cm
\hspace*{0.2cm}\epsffile{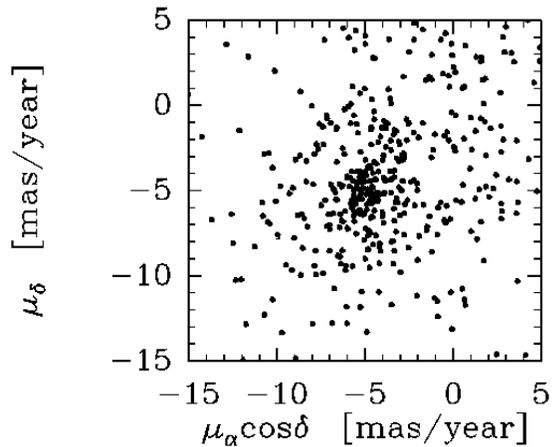}

\caption[]{The vector-point-plot diagram of all stars in the
field of M10}
\label{f_vppdia}
\end{figure}

\section {Separation of cluster and field stars}

Figure~\ref{f_vppdia}  gives the vector-point-plot diagram of all our proper 
motions indicating the concentration of the  cluster stars.
For the separation of cluster and field stars we have concentrated
our investigation on stars up to the limiting radius of 
14$\arcmin$ (Webbink 1988) around the cluster centre.
284 stars remained in the data sample.
For these stars a bivariate Gaussian fitting to the proper motions 
was performed using the method of Sanders (1971).
Table~\ref{t_fitpar}  gives the result of the analysis.
The cluster standard deviations are larger than the internal errors of the 
proper motion.
This difference is clearly not an indication of a possible detection
of internal motions of stars in the cluster. 
More likely these differences are caused by crowding effects and undetected
systematic errors of the different telescopes used in this investigation.

M10 is a globular cluster with three slow variable stars (Clement et al.\, 1985)
and several UV bright stars (Zinn et al.\, 1972; Harris et al.\, 1983).
Our sample contains only the variable stars V2 and V3 (see references in 
Clement et al.\, 1985).
The other special stars are located in crowded regions,
below the limiting magnitude, 
or exhibit too large proper motion errors.
Both variable stars V2, V3 have according to our investigation a membership
probability of 97\%.  
This confirms the earlier suggestion (Clement et al.\, 1985) that
these stars are of W-Virginis type,
although V3 has a very unusual period of 7.8~days.

\begin{table}
\caption[]{Parameters of the fit to the bivariate Gaussian 
distribution of the proper motions of stars with $R<14\arcmin$ 
around M10.  
$n_{\rm c}$ and $n_{\rm f}$  denote the number of the cluster and field stars,
$x_{\rm c}$, $y_{\rm c}$ the mean cluster proper motion in $\alpha$ and
$\delta$, respectively, $x_{\rm f}$, $y_{\rm f}$ the mean proper motion
of field stars, $\sigma_{x{\rm c}}$, $\sigma_{y{\rm c}}$ the Gaussian widths of
the cluster distribution, and $\sigma_{x{\rm f}}$, $\sigma_{y{\rm f}}$ the
widths of the field star distribution.
The $x$-direction is the direction of right ascension, the $y$ direction
is the declination axis.
The units are mas/yr}
\begin{flushleft} 
\begin{tabular}{lrlr}
\hline\noalign{\smallskip}
Parameter & Value & Parameter & Value \\  
\noalign{\smallskip}\hline\noalign{\smallskip}
$n_{\rm c}$    & 193  & $n_{\rm f}$    & 88  \\ 
$x_{\rm c}$    & $-4.9$ & $x_{\rm f}$    & $-3.5$ \\
$y_{\rm c}$    & $-5.3$ &  $y_{\rm f}$    & $-3.3$ \\
$\sigma_{x{\rm c}}$ & 1.8 & $\sigma_{x{\rm f}}$ & 7.0  \\
$\sigma_{y{\rm c}}$ & 2.0 & $\sigma_{y{\rm f}}$ & 7.1  \\
\noalign{\smallskip}\hline
\end{tabular}
\end{flushleft}
\label{t_fitpar}
\end{table}

\section {The space motion of M10}

\begin{figure} 
\epsfysize=6.0cm
\hspace*{0.6cm}\epsffile{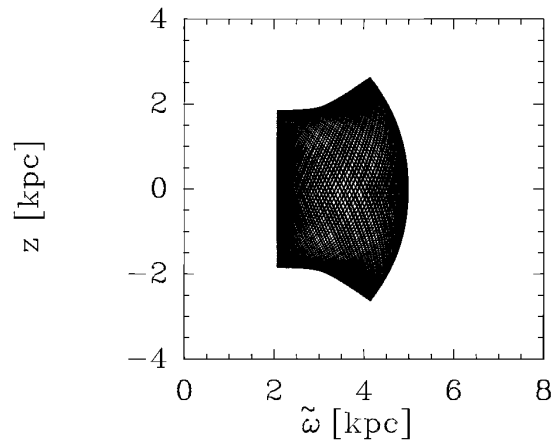}
\caption[]{The meridional section of the orbit of M10 calculated
backwards over 10~Gyr}
\label{f_m10orb}
\end{figure}

The first absolute  proper motion of M10 based on the Hipparcos system
was given in Geffert et al.\, (1997),
which was later used in Odenkirchen et al.\, (1997) and
Dinescu et al.\, (1999). 
However, as mentioned in Odenkirchen et al.\, (1997),
the result of M10 was only preliminary due to the small number of
Hipparcos stars used for the reduction of the plates.
The use of additional plates from
Shanghai has led to a significant lower proper motion in declination.
However, even for this sample of plates the solution is
only moderately stable: Omitting one reference star 
leads  to an absolute proper motion of M10, which  
differs from the original solution by about more than the mean
uncertainties of the absolute proper motions. 
Fortunately, this is not the case  for the denser ACT catalogue. 
In summary, the use of the ACT catalogue 
resulted in a more stable solution, while the use of Hipparcos 
stars provided the more direct link to an extragalactic
reference frame.
The catalogues based on the ACT catalogue (Urban et al.\, 1998) and 
on the Hipparcos catalogue (ESA 1997) contributed 
therefore with equal
weight to the final absolute proper motion of the globular
cluster M10.
We obtained a mean absolute proper motion of $\mua = -5.5 \plm 1.2\;$mas/yr
and $\mud = -6.2 \plm 1.2\;$mas/yr.
Together with the distance from the Sun of 4.3~kpc and a radial velocity of
$+76\;$\kms (Harris 1996) we have calculated the velocity components in
a system of galactic standard at rest ($U$,$V$,$W$), peri-- and apogalactic
distances $R_{\rm p}$, $R_{\rm a}$, eccentricity $e$ and the $z$-component of
the angular momentum of the orbit using a simple logarithmic galactic mass
model (see e.g. Dauphole et al.\, 1996).
These data are shown in Table~\ref{t_m10dat}.
Figure~\ref{f_m10orb}  gives the orbit of M10 integrated over 10~Gyr backwards
according to the method of Allen \& Santill\'an (1991) using
the programme of Odenkirchen et al.\, (1997).

\begin{table*}
\caption[]{Velocity components in a system of galactic standard at rest and
           kinematical data of the orbit of globular cluster M10.
           A simple logarithmic galactic mass model was used
           (see e.g. Dauphole et al. 1996).
           (Note that $U$ points from Sun to galactic centre!
            A positive $z$ component of the angular momentum indicates
            a prograde rotation in the Galaxy.)}
\begin{flushleft}
\begin{tabular}{lrrrrrrr}
\hline\noalign{\smallskip}
  Object  & \multicolumn{1}{c}{$U$} & \multicolumn{1}{c}{$V$} &
  \multicolumn{1}{c}{$W$} & \multicolumn{1}{c}{$R_{\rm a}$} &
  \multicolumn{1}{c}{$R_{\rm p}$} & \multicolumn{1}{c}{$e$} &
  \multicolumn{1}{c}{$I_{z}$} \\
  & \multicolumn{1}{c}{[km~s$^{-1}$]} & \multicolumn{1}{c}{[km~s$^{-1}$]} &
  \multicolumn{1}{c}{[km~s$^{-1}$]} & \multicolumn{1}{c}{[kpc]} & 
  \multicolumn{1}{c}{[kpc]} & & \multicolumn{1}{c}{[kpc km~s$^{-1}$]} \\
\noalign{\smallskip}
\hline\noalign{\smallskip}
M10 & $+111 \pm 10$ & $+90 \pm 18$ & $+63 \pm 18$ & $5.4 \pm 0.2$ &
 $2.3 \pm 0.4$ & $0.41 \plm 0.08$ & $+533 \plm 70$  \\
\noalign{\smallskip}
\hline
\end{tabular}
\end{flushleft}
\label{t_m10dat}
\end{table*}

\section{Discussion}

Our results demonstrate the importance  of the new astrometric catalogues
(Hipparcos, ACT) for the determination of absolute proper motions of objects
located in the galactic plane.
For these objects no extragalactic link of the proper motion system
is available due to the absense of suitable objects.
We show that the differences between the solutions based on the ACT catalogue
and the corresponding one based on the Hipparcos catalogue are small.
Although this result has to be checked using data of other fields,
it would offer a range of applications for the ACT catalogue
because of its larger number of stars.
Moreover, even reaching only moderate accuracy on the wide field CCD frames 
with respect to small field CCD observations, our results indicate that
these observations may supersede the photographic plates in near future. 

The new data of M10 show an eccentricity of the orbit of 0.41,
which would place the cluster now in the range of eccentricities
from 0.4 to 0.8 found for the majority of the globular clusters
(Odenkirchen et al.\, 1997).
The rotational velocity changed to $\Theta = +111 \plm 17\;$\kms\ with respect 
to the $+144\;$\kms\ found in Odenkirchen et al. (1997). 
According to its moderate eccentricity and rotational velocity,
M10 belongs rather to the halo class of objects.
This is in line with its metallicity.
However, the $z$-distance from the galactic plane does not exceed 3~kpc,
which would be more characteristic for a thick disk object.

As noted already in Dinescu et al.\, (1999) recent age determinations
(Hurley et al.\, 1989; Richer et al.\, 1996; Chaboyer et al.\, 1996;
Buonano et al.\, 1998) agree that M10 is an old halo cluster.
Together with NGC~6626, NGC~6752 (Dinescu et al.\, 1999) and M71,
which has according to Geffert \& Maintz (2000) a much higher age,
M10 establishes a group of high age globular clusters with orbits of
a disk or thick disk character.
This result would imply that at the time when the oldest globular clusters
were born, the Milky Way had already a gas distribution with a disk component,
where the gas enabled the formation of clusters.
On the other hand the recent age determination by Rosenberg et al.\, (1999)
places M10 in the middle of the age distribution of globular clusters. 

Piotto \& Zoccali (1999) have found an unusual steep luminosity
function of M10 in comparison with other clusters.
Our data indicate that this luminosity function was generated 
more likely  by internal dynamics of the cluster than by the motion 
in the Galaxy. 
If the luminosity function would be affected by the crossing of the cluster
through the galactic plane, one would rather expect a more flat luminosity
function for M10 due to the evaporation of faint stars as it was found
recently for the globular cluster NGC~6712 (De Marchi et al.\, 1999). 

\acknowledgements 
We are indebted to H.-J. Tucholke (Bonn) for measuring some of
the refractor plates.
It is a pleasure to thank K.S.~de~Boer (Bonn) for helpful discussions.
This research has made use of the Simbad database (see Wenger et al. 2000),
operated at CDS, Strasbourg, France.

{}

\end{document}